\documentclass[%
a4paper,							% paper format
11pt,								% Schriftgröße (12pt, 11pt (Standard))
bibliography=totoc,						% Literaturverzeichnis im Inhalt
abstracton,					% Überschrift über der Zusammenfassung an	
]
{scrartcl}
\usepackage[a4paper,left=3.0cm,right=2.5cm, top=2.5cm, bottom=3.0cm]{geometry}
\usepackage[headsepline=.4pt,footsepline=.4pt,automark,autooneside=false,]{scrlayer-scrpage}
\clearpairofpagestyles
\automark[subsection]{section} 
\automark*[subsubsection]{subsection}
\ihead{\scriptsize\rightmark} 
\usepackage[ngerman, english]{babel}
\usepackage[T1]{fontenc}
\usepackage[utf8]{inputenc}
\usepackage{lmodern} %latin modern

\usepackage{calc}
\usepackage{xspace}

\usepackage{amsmath}
\usepackage{amssymb}
\usepackage{amsfonts}
\usepackage{amsthm,thmtools}
\usepackage{mathtools}
\usepackage{bbm}

\theoremstyle{plain}

\theoremstyle{definition}

\theoremstyle{remark}

\usepackage{graphicx,float}
\usepackage[format=plain]{caption}

\usepackage{setspace}

\usepackage{xcolor}

\usepackage{fancyvrb}

\usepackage{paralist}

\usepackage{tabularx}
\usepackage{booktabs}
\usepackage{multirow}
\usepackage{multicol}
\usepackage{pdflscape}
\usepackage{diagbox}

\usepackage{todonotes}
\usepackage[final]{showkeys}

\usepackage[
	backend=bibtex,
	style=authoryear-comp,
	maxbibnames=9,
	maxcitenames=2,
	dashed=false,
	natbib=true,
	sortcites=true,
	block=space
]{biblatex}

\usepackage{orcidlink}
\newcommand{\orcid}[1]{\textsuperscript{\orcidlink{#1}}}
\usepackage{titleref}
\usepackage{hyperref}
\usepackage{cleveref}
\hypersetup{
    colorlinks,
    linkcolor={red},
    citecolor={blue},
    urlcolor={blue}
}

\newenvironment{key}[1][]{ {\noindent \bf Keywords#1: } \rmfamily
}{\hspace*{\fill}  \\}

\newenvironment{code}[1][]{ {\noindent \bf Code availability#1: } \rmfamily
}{\hspace*{\fill}  \\}%

\renewcommand{\P}{\mathbb{P}}
\newcommand{\Q}{\mathbb{Q}}
\newcommand{\E}{\mathbb{E}}
\newcommand{\R}{\mathbb{R}}
\newcommand{\N}{\mathbb{N}}

\newcommand{\currency}{NOK}
\newcommand{\conversionRate}{\mathrm{c}}
\newcommand{\FC}{\mathrm{F}}
\newcommand{\cumFC}{\mathrm{CF}}
\newcommand{\HC}{\mathrm{h}}
\newcommand{\cumHC}{\mathrm{CH}}
\newcommand{\numFishes}{\mathrm{H}}
\newcommand{\moratlityRate}{\mathrm{m}}
\newcommand{\biomass}{\mathrm{B}}
\newcommand{\growth}{\mathrm{w}}
\newcommand{\PC}{\mathrm{PC}}
\newcommand{\liceThreshold}{l}
\newcommand{\LPF}{\mathrm{LPF}}
\newcommand{\cumRemovals}{\mathrm{R}}
\newcommand{\BC}{\mathrm{BC}} %total biological costs
\newcommand{\treatmentCosts}{\mathrm{tc}} % percentage of treatment costs
\newcommand{\cumTreatmentCosts}{\mathrm{Tc}} % cumulative treatment costs

\newcommand{\relImprovement}{\mathrm{RI}}
\newcommand{\fromDate}{01/01/2012\xspace}
\newcommand{\toDate}{01/06/2023\xspace}

\newcommand{\CPU}{Intel(R) Core(TM) i9-13900K CPU @ 3.00\,GHz\xspace}
\newcommand{\RAM}{2x32\,GB (Dual Channel) Kingston DIMM DDR5 RAM @ 5600 MHz\xspace}
\newcommand{\GPU}{NVIDIA GeForce RTX 4090 (24\,GB GDDR6X RAM)\xspace}
\newcommand{\OS}{Windows 11 Pro\xspace}

\SaveVerb{python}=(Intel-)Python 3.10=
\newcommand{\python}{\protect\UseVerb{python}\xspace}
\SaveVerb{tensorflow}=Tensorflow 2.8.0=
\newcommand{\tensorflow}{\protect\UseVerb{tensorflow}\xspace}
\SaveVerb{softmax}=softmax=

\SaveVerb{relu}=relu=

\SaveVerb{matlab}=Matlab 2023a=
\newcommand{\matlab}{\protect\UseVerb{matlab}\xspace}
\SaveVerb{ga}=ga=

\SaveVerb{fmincon}=fmincon=
\newcommand{\fmincon}{\protect\UseVerb{fmincon}\xspace}
\SaveVerb{fitdist}=fitdist=

\SaveVerb{lsqnonlin}=lsqnonlin=
\newcommand{\lsqnonlin}{\protect\UseVerb{lsqnonlin}\xspace}
\newcommand{\matlabGOtoolbox}{(Global) Optimization Toolbox\xspace}

\begin{document}
\onehalfspacing
\pagenumbering{Roman}
\begin{titlepage}
\vspace*{.33cm}
\begin{center}
    \huge
    \bfseries 
    \sffamily
    On the Impact of Biological Risk in Aquaculture Valuation and Decision Making
\end{center}
\renewcommand{\thefootnote}{\fnsymbol{footnote}}
\footnotetext[1]{Department of Mathematics and Statistics, Umeå University, Sweden}
\footnotetext[2]{Inland University of Applied Sciences, Lillehammer, Norway}
\footnotetext[3]{E-mail: christian.ewald@umu.se and kevin.kamm@umu.se}

\begin{center}
    \Large
    Christian-Oliver Ewald%
    \footnotemark[1]{}\textsuperscript{,}%
    \footnotemark[2]{}\textsuperscript{,}%
    \footnotemark[3]{}\textsuperscript{,}%
    \orcid{0000-0003-3288-0164}%
    \hspace{2em}
    Kevin Kamm%
    \footnotemark[1]{}\textsuperscript{,}%
    \footnotemark[3]{}\textsuperscript{,}%
    \orcid{0000-0003-2881-0905}
\end{center}
\begin{center}
    \Large
    \today
\end{center}
\vspace*{.5cm}

\renewcommand{\thefootnote}{\arabic{footnote}}

% \title{On the Impact of Parasite Risk in Aquaculture Valuation and Decision Making}
% \author{Christian-Oliver Ewald\thanks{Department of Mathematics and Statistics, Umeå University, Sweden, and Inland University of Applied Sciences, Lillehammer, Norway, e-mail: christian.ewald@umu.se}\orcid{0000-0003-3288-0164} \and Kevin Kamm\thanks{Department of Mathematics and Statistics, Umeå University, Sweden, e-mail: kevin.kamm@umu.se}\orcid{0000-0003-2881-0905}}
% \maketitle 

\begin{abstract}
This paper explores the impact of stochastic mortality and disease on animal-based commodities, with a specific emphasis on aquaculture, particularly in the context of salmon farming. The investigation delves into the stochastic nature of mortality and treatment plans based on historical data related to salmon lice. Given that salmon lice pose a significant challenge in salmon farming, with associated treatment costs estimated to be comparable to feeding expenses, their removal is imperative to ensure the survival of the salmon and comply with the Norwegian government's stipulation of maintaining 0.5 lice per fish.

We propose a new model that considers the relationship between hosts and parasites to estimate the number of treatments required and their overall cost. An important aspect of this model is its incorporation of stochastic effectiveness for each removal. After calibrating the model to the available data, the study examines the stochastic behavior's impact on the optimal harvesting decision in comparison to deterministic mortality models. The results indicate an approximate $1.5\,\%$ increase in the value of the salmon farm when employing the harvesting rule based on the stochastic host-parasite model as opposed to a deterministic model.
\end{abstract}

\medskip
\begin{key}
Deep Learning, Optimal Stopping, Real Options, Host-Parasite Model, Aquaculture
\end{key}

% \medskip

%\begin{ams}
%91B28
%\end{ams}

% \begin{jel}
% C61, E21, O41
% \end{jel}

\medskip
\begin{code}
The Matlab/ Python code and data sets to produce the numerical experiments are available at
\url{https://github.com/kevinkamm/AquacultureStochasticMortality}.
\end{code}

\thispagestyle{empty}
\end{titlepage}

\newpage
\pagestyle{scrheadings}\ihead{\scriptsize\rightmark}\pagenumbering{arabic}

%%%%%%%%%%%%%%%%%%%%%%%%%%%%%%%%%%%%%%%%%%%%%%%%%%%%%%%%%%%%%%%
%%% 
%%%%%%%%%%%%%%%%%%%%%%%%%%%%%%%%%%%%%%%%%%%%%%%%%%%%%%%%%%%%%%%
\section{Introduction}\label{sec:introduction}

This study draws parallels with research conducted by \citet{EK2023}, which explored the influence of stochastic feeding costs on optimal decision-making in aquaculture management. However, we will focus in this article on the effects of stochastic mortality instead of feeding. Our commodity models align with those outlined in \citet{EK2023}.

Specifically, we examine stochastic mortality models within the context of salmon farms. We do not focus on the intrinsic Salmon mortality, but rather the mortality due to parasites known as salmon lice. These parasites attach themselves to the salmon's skin and feed on its blood. If left unchecked, infestation can lead to mortality, which can harm neighboring fish farms. To prevent adverse effects, the Norwegian government has set a threshold of 0.5 female lice per fish. As a result, farmers must take countermeasures to stay below this threshold.

Salmon lice treatments can be classified into three main categories: medical, biological, and mechanical. It is worth noting that biological treatments have been phased out due to ethical concerns surrounding the sacrifice of other fish species. Currently, the predominant method involves mechanical removal, which includes techniques such as warm or freshwater baths, brushing, or laser methods. Even though very small, these methods pose some risks to salmon as well, which are carefully considered in this study.

Accurate modeling of treatment times and associated costs is crucial to optimize the value of a fish farm and determine the optimal harvesting time. According to \citet{Misund2022}, biological costs are now as significant as feeding costs, highlighting the need for a comprehensive examination of stochastic mortality.

Our contribution involves introducing a novel model for treatment times and their impact on salmon mortality, rooted in the biological dynamics of the problem. Drawing inspiration from \citet{AndersonMay1978}, we employ a host-parasite model. We demonstrate the calibration of this model to data and assess the impact of stochasticity on its deterministic counterpart, represented by the mean of the model. The results suggest a $1.5\,\%$ increase in farm value when using a stochastic model.

%%%%%%%%%%%%%%%%%%%%%%%%%%%%%%%%%%%%%%%%%%%%%%%%%%%%%%%%%%%%%%%%%%%%%
%%%
%%%%%%%%%%%%%%%%%%%%%%%%%%%%%%%%%%%%%%%%%%%%%%%%%%%%%%%%%%%%%%%%%%%%%%
\paragraph{Literature review.}
For an overview of the historical development of salmon commodity pricing and valuation models, we refer the reader to \citet{Ewald2017}. Additionally, for a comprehensive treatment of all the economic factors of fish farming, we refer the reader to \citet{Misund2022} and \citet{Luna2023}. For a treatment of stochastic feeding costs, we refer the reader to \citet{EK2023}.
In this review, we focus on (stochastic) mortality.

Deterministic host-parasite models have been extensively explored in biological contexts, with notable studies conducted by \citet{AndersonMay1978}, \citet{Milner1999}, \citet{Tadiri2019}, among others. In particular, \citet{Frazer2012} considered a deterministic model specifically for salmon lice, which is related to the model considered in this paper. In addition, \citet{Kragesteen2019} have examined lice thresholds for both individual farms and farm networks to determine the level of lice per fish that warrants treatment. They used a deterministic model of parasite growth with a simplified harvesting rule. In addition, \citet{Asplin2020} and \citet{Sandvik2020} explored sophisticated models that describe in detail the spread of lice taking hydrodynamic effects into account.

To the best of our knowledge, this article is the first to study harvesting rules that depend on stochastic mortality caused by salmon lice.
%%%%%%%%%%%%%%%%%%%%%%%%%%%%%%%%%%%%%%%%%%%%%%%%%%%%%%%%%%%%%%%%%%%%%
%%%
%%%%%%%%%%%%%%%%%%%%%%%%%%%%%%%%%%%%%%%%%%%%%%%%%%%%%%%%%%%%%%%%%%%%%%

\medskip
The remainder of this paper is structured as follows:
In \Cref{sec:framework}, we introduce the commodity models, followed in \Cref{sec:salmonFarmParameters} by a description of the salmon farm features considered in this paper. The mortality model will be introduced in \Cref{sec:stochMortality}.
Afterwards, in \Cref{sec:stopping}, we define our optimal stopping problem using both deterministic and stochastic feeding costs.
The numerical results will be discussed in \Cref{sec:numerics}, which first explains the market data in \Cref{sec:marketData} and salmon lice data in \Cref{sec:liceData}, followed by a calibration algorithm in \Cref{sec:calibration} for the model. Then, we explain our methodology for comparing stochastic and deterministic feeding rules in \Cref{sec:stoppingBoundaries} and reach our conclusion in \Cref{sec:conclusion}.

%%%%%%%%%%%%%%%%%%%%%%%%%%%%%%%%%%%%%%%%%%%%%%%%%%%%%%%%%%%%%%%%%%%%%%%%%%%%%%%%%
%%% 
%%%%%%%%%%%%%%%%%%%%%%%%%%%%%%%%%%%%%%%%%%%%%%%%%%%%%%%%%%%%%%%%%%%%%%%%%%%%%%%%%
\section{Mathematical and Model Framework}\label{sec:framework}
Following \citet{EK2023}, let henceforth $\left(\Omega,\mathcal{F},\Q\right)$ be a probability space and $\Q$ be a risk-neutral measure. Moreover, let $T>0$ be a finite time horizon and $r>0$ a fixed deterministic interest rate.
We will use a multi-commodity framework consisting of two independent Schwartz-2-factor models directly under $\Q$, i.e.
\begin{align*}
    dS_t^i &= \left(r-\delta_t^i\right) S_t^i dt + \sigma^i_1 S_t^i dW_t^{i,1}, \quad S_0^i=s_0^i \in \R_{\geq 0,}\\
    d\delta_t^i &= \left(\kappa^i\left(\alpha^i -\delta^i_t\right) -\lambda^i\right) dt + \sigma_2^i dW_t^{i,2},  \quad \delta_0^i=d_0^i \in \R\\
    d\langle W^{i,1},W^{i,2}\rangle_t&= \rho^i dt,\qquad d\langle W^{i,1},W^{j,2}\rangle_t=0,\quad i\neq j,
\end{align*}
where $W_t\coloneqq \left(W_t^{i,j}\right)_{i,j=1,2}$ are Brownian motions generating the $\sigma$-algebra $\mathcal{F}_t$ augmented by $\Q$ nullsets satisfying the usual conditions.

The dynamics $dS_t^i$ describe the i-th commodity's spot price with convenience yield described by $d\delta_t^i$. The parameter $\sigma_1^i>0$ is the spot volatility, $\kappa^i>0$
the mean reversion speed of the convenience yield, $\alpha^i\in \R$ the long-term mean, $\lambda^i\geq 0$ a risk-premium and $\sigma_2^i>0$ the volatility of the convenience yield. For a treatment of future prices in this framework we refer the reader to \citet{EK2023} since this will not be the main focus of this article.

%%%%%%%%%%%%%%%%%%%%%%%%%%%%%%%%%%%%%%%%%%%%%%%%%%%%%%%%%%%%%%%%%%%%%%%%%%%%%%%%%
%%% 
%%%%%%%%%%%%%%%%%%%%%%%%%%%%%%%%%%%%%%%%%%%%%%%%%%%%%%%%%%%%%%%%%%%%%%%%%%%%%%%%%
\subsection{Salmon Farm Parameters}\label{sec:salmonFarmParameters}
In this subsection, we briefly describe the features of a salmon farm, which we consider in this paper, and refer the reader to \citet{Misund2022} for a detailed economical treatment of aquaculture farms.

We will assume a single rotation fish farm, which means that we assume that a fish farmer buys $\numFishes_0\in \N$ smolt, young salmon, and feeds them till they are ready for harvest, and after the harvest, our consideration is finished. 
The growth of the fish over time depends on various factors like water temperature, amount and quality of feed, health, etc. We simplify this by considering a deterministic function over time called \emph{Bertalanffy’s growth function}, which is given by 
\begin{align*}
    \growth_t &\coloneqq \growth_\infty \left(a-b\, e^{-ct}\right)^3,\\
    % \left(\partial_t \growth\right)(t)&= 3\growth_\infty \left(a-b\, e^{-ct}\right)^2\, b\,c\, e^{-ct},
\end{align*}
for some parameters $a,b,c$ given in \Cref{tab:fishPoolParameters}. This function measures the growth in kg per fish. 
Hence, we need to keep track of our initial population $n_0$ over time as well.

%%%%%%%%%%%%%%%%%%%%%%%%%%%%%%%%%%%%%%%%%%%%%%%%%%%%%%%%%%%%%%%%%%%%%%%%%%%%%%%%%
%%% 
%%%%%%%%%%%%%%%%%%%%%%%%%%%%%%%%%%%%%%%%%%%%%%%%%%%%%%%%%%%%%%%%%%%%%%%%%%%%%%%%%
\subsubsection{Stochastic Mortality and Treatment Costs}\label{sec:stochMortality}
In our previous article \citet{EK2023}, we assumed a fixed rate of mortality $\moratlityRate>0$. The total number of fishes $\numFishes_t$ with mortality rate $\moratlityRate$ was modeled as another deterministic function
\begin{align*}
    d \numFishes_t &= -\moratlityRate\, \numFishes_t dt, \quad \numFishes_0>0.
\end{align*}
Then, we can measure the total biomass (in kg) of the fish farm over time by setting
\begin{align*}
    \biomass_t = \numFishes_t \growth_t.
\end{align*}

Now, we want to extend this model by using a more sophisticated model for $\numFishes_t$ and meanwhile model the number of mechanical lice removals, such that the threshold of $\liceThreshold=0.5$ female lice per fish is not exceeded. It is straightforward to include time-varying thresholds $\liceThreshold$ and indeed they sometimes vary between $0.2$ and $0.5$ depending on the region and time of the year. 
%%%%%%%%%%%%%%%%%%%%%%%%%%%%%%%%%%%%%%%%%%%%%%%%%%%%%%%%%%%%%%%%%%%%%%%%%%%%%%%%%
%%%
%%%%%%%%%%%%%%%%%%%%%%%%%%%%%%%%%%%%%%%%%%%%%%%%%%%%%%%%%%%%%%%%%%%%%%%%%%%%%%%%%
\paragraph*{Host-parasite model.}
Let us denote by $H_t$ the host population and $P_t$ the parasite population. A notable difference to standard SIR models (Susceptible-Infectious-Recovered) is that we can allow an infected species to be infected again, i.e., have more than one louse. Therefore, we need to model the number of lice and fish as a system of coupled dynamics. Since the lice threshold shall not be exceeded, otherwise extermination of the farm would occur under some circumstances, we will include the threshold explicitly into the dynamics. We only model mechanical removals. They have a large success rate of removals but are usually only performed on a part of the farm instead of the whole site. Thus, many removals are usually needed during the growing period and each removal introduces some cost. During each removal, we will assume a random success rate of lice removal and a small random number of salmon not surviving the treatment, e.g., hot water baths can be stressful for the salmon as well, mechanical brushes can hurt the salmon, etc.

Let $0\leq t_1 \leq t_2 \leq \dots$ be the times of the treatments, i.e. the times such that 
$\frac{P_t}{H_t} \geq\liceThreshold_t$, then we define our host-parasite model for $t< t_1$ as for generic processes $h_t$ and $p_t$ determined below
\begin{align*}
    dH_t &= h_t(P_t,H_t) dt, \quad H_0>0\\
    dP_t &= p_t(P_t,H_t) dt, \quad P_0>0
\end{align*}
then restart at for $t\in [t_1,t_2)$
\begin{align*}
    dH_t &= h_t(P_t,H_t) dt, \quad H_{t_1}=X_{t_1}\, H_{t_1-}\\
    dP_t &= p_t(P_t,H_t) dt, \quad P_{t_1}=Y_{t_1}\, P_{t_1-}
\end{align*}
and so on.

Since we are no experts in biology, especially salmon lice, we restrict ourselves to a simple model not taking the different stages of parasites in their life-cycle into account. We encourage our readers to modify the model according to their requirements and test it using the publicly available code.
To keep our experiments simple, we have used deterministic dynamics for the lice, which can be estimated from biological experiments. Although the growth of parasites and salmon depends on various factors such as water temperature, salinity, and other random factors, we have assumed that the farmer has a good understanding of these parameters for their farm, and hence, we have excluded them from our dynamics. In particular, we chose exponential growth models based on \citet{AndersonMay1978}
\begin{align}
    \begin{aligned}[c]
    h_t(P_t,H_t)&=-\left(\mu + \alpha \frac{P_t}{H_t}\right) H_t,&& X_t \sim \mathrm{unif}(0.995,1)\\
    p_t(P_t,H_t)&=\left(\lambda \frac{H_t}{H_0} - (b+\mu) -\alpha \frac{P_t}{H_t}\right) P_t,&& Y_t\coloneqq 0.1 + 0.8\tilde{Y}_t,\ \tilde{Y}_t \sim \mathrm{beta}(\beta_1,\beta_2)
    \end{aligned}
    \label{eq:hostParasite}
\end{align}

We chose $Y_t$ in this way to ensure a minimal success rate of $10\,\%$ and a maximal success rate of $90\,\%$, which sounded more realistic and avoided pathological examples with zero and complete success rates.
Moreover, we will denote by $N_t$ the counting process for the necessary mechanical removals per path, i.e., the hitting times 
$\frac{P_t}{H_t}=\liceThreshold(t)$, which is well defined, since the paths of $H_t,P_t$ prior to the hitting boundary are continuous.

The parameter $\mu$ controls the intrinsic mortality of salmon and will be fixed to $5\,\%$ for our tests and similarly 
$b$ controls the intrinsic mortality of salmon lice and will also be fixed to $5\,\%$. The salmon mortality due to salmon lice can be adjusted by $\alpha$ and increases multiplicatively with the ratio of lice per fish. We will fix it to $10\,\%$ in our experiments, which might seem high but we will see that the impact of this factor is small as long as the lice per fish is kept small as well, which is the purpose of the treatments. The salmon population cannot grow over time, so there are no births, but the lice population will increase exponentially if left unchecked. This rate of population growth $\lambda$ depends in this model on the relative number of salmon in the farm and will be calibrated from the data alongside the effectiveness of lice removals by finding suitable parameters $\beta_1$ and $\beta_2$ for the beta-distribution.

We tried other distributions as well, but the beta distribution matched the data best. Also, the mortality due to treatments is very sensitive to the lower bound of the uniform random number. Changing the lower bound to $0.99$ leads to an overestimation of mortality.

%%%%%%%%%%%%%%%%%%%%%%%%%%%%%%%%%%%%%%%%%%%%%%%%%%%%%%%%%%%%%%%%%%%%%%%%%%%%%%%%%
%%%
%%%%%%%%%%%%%%%%%%%%%%%%%%%%%%%%%%%%%%%%%%%%%%%%%%%%%%%%%%%%%%%%%%%%%%%%%%%%%%%%%
\paragraph{Treatment costs.}
We will assume for simplicity that each treatment will remove a certain percentage from the current value of the salmon farm, i.e. the total treatment costs are given by
\begin{align*}
    \cumTreatmentCosts_t =  \treatmentCosts \sum_{0\leq s \leq t}{\Delta H_s},\quad \Delta H_s\coloneqq H_s - H_{s-}, \quad \treatmentCosts\in [0,1].
\end{align*}
The treatment times are identified by the jumps in the host population process $\Delta H_s$ and are in the host-parasite model the times, where the lice-threshold $\liceThreshold$ is exceeded.

% %%%%%%%%%%%%%%%%%%%%%%%%%%%%%%%%%%%%%%%%%%%%%%%%%%%%%%%%%%%%%%%%%%%%%%%%%%%%%%%%%
% %%%
% %%%%%%%%%%%%%%%%%%%%%%%%%%%%%%%%%%%%%%%%%%%%%%%%%%%%%%%%%%%%%%%%%%%%%%%%%%%%%%%%%
% \paragraph{Average models.}
We want to compare the impact of the stochasticity to the harvesting decision rule. For this, we want to compare the harvesting decisions using stochastic and deterministic mortality. Similar to \citet{EK2023}, to make these models comparable, we derive the deterministic models by taking the mean of the stochastic model, i.e.
\begin{align*}
    H_t^{\text{determ}}\coloneqq \E\left[H_t\right], \quad 
    \cumTreatmentCosts_t^\text{determ}\coloneqq \E\left[\cumTreatmentCosts_t\right].
\end{align*}

This also implies that the biomass $\biomass_t = \numFishes_t \growth_t$ in this model will either be stochastic or deterministic depending on the choice of $\numFishes_t$.
%%%%%%%%%%%%%%%%%%%%%%%%%%%%%%%%%%%%%%%%%%%%%%%%%%%%%%%%%%%%%%%%%%%%%%%%%%%%%%%%%
%%% 
%%%%%%%%%%%%%%%%%%%%%%%%%%%%%%%%%%%%%%%%%%%%%%%%%%%%%%%%%%%%%%%%%%%%%%%%%%%%%%%%%
\subsubsection{Feeding and Harvesting Costs}\label{sec:feedingCosts}
We will consider only two varying factors of production costs: harvesting costs and feeding costs in this article, all the other costs like labor, medical treatments, capital costs, etc., will be treated as constants for simplicity. 
The harvesting costs are given by $\HC_0$ per kg of fish and the total harvesting cost of the fish farm at time $t>0$ is therefore
\begin{align*}
    \cumHC_t=\HC_0\, \biomass_t.    
\end{align*}
For this, we set a conversion rate of how much kg of feed will convert to kg of fish to
\begin{align*}
    \conversionRate \coloneqq 
    1.1\frac{\text{\,kg feed}}{\text{\,kg fish}}.
\end{align*}
Let $F(0)=F_0$ be a given feeding cost for one fish per year. We will infer the feeding costs from the relative changes of soybean prices $S_t^2$, i.e.
\begin{align*}
    \tilde{S}_t^2 \coloneqq \frac{S_t^2}{S_0^2},
\end{align*}
by using
\begin{align*}
    \FC^\text{stoch}_t = \FC_0 \tilde{S}_t^2, \quad \FC^\text{determ}_t = \FC_0 \E\left[\tilde{S}_t^2\right],
\end{align*}
and define the discounted cumulative total feeding costs as 
\begin{align*}
    \cumFC_t = \int_0^t{ e^{-rs} \left(\FC_s\, \numFishes_s \left(\partial_t \growth\right)(s)\, \conversionRate\right)  ds}
\end{align*}
for general feeding costs $\FC_t$.

The parameters in \Cref{tab:fishPoolParameters} are mostly taken from \citet{Ewald2017} and references therein. The initial feeding, harvesting, and biological costs are estimated from \citet[p.~25 Figure 9]{Misund2022}. We will use the values in \Cref{tab:fishPoolParameters} for the remainder of this paper.
\begin{table}[]
    \centering
    \caption{Fish farm parameters.}
    \begin{tabular}{l*{3}{c}}
        Name                        & Symbol            & Value                                 & Unit\\
        \toprule
        Bertalanffy growth factor   & $a$               & 1.113                                 &  -     \\    
        Bertalanffy growth factor   & $b$               & 1.097                                 &  -     \\ 
        Bertalanffy growth factor   & $c$               & 1.43                                  &  -     \\ 
        Asymptotic weight           & $\growth_\infty$  & 6                                     & kg    \\ 
        Mortality rate              & $\moratlityRate$  & 5                                     & \%    \\
        Conversion rate             & $\conversionRate$ & 1.1                                   & kg feed/ kg fish\\
        Number of recruits          & $\numFishes_0$    & 10000                                 & fish \\
        Time horizon                & $T$               & 3                                     & years\\
        Harvesting possibilities    & $N$               & 72                                    & -\\
        Salmon spot price           & $\hat{S}_0^1$     & see \eqref{eq:soySalmonParam}         & \currency/ kg\\
        Production costs            & $\PC$             & $0.5\, \hat{S}_0^1$                   & \currency/ kg\\
        Harvesting costs            & $\HC_0$           & $0.1\, \PC$                           & \currency / kg\\
        Feeding costs               & $\FC_0$           & $0.25\, \PC$                          & \currency / kg year\\
        Biological costs            & $\BC_0$           & $0.3\, \PC$                           & -\\
        Salmon initial value        & $S_0^1$           & $\hat{S}_0^1 -\PC + \HC_0 + \FC_0 +\BC_0$    & \currency/ kg\\
        Soy initial value           & $S_0^2$           & 1                                     & - \\
        Treatment costs             & $\treatmentCosts$ & 1.5                                   & \% \\
    \end{tabular}
    \label{tab:fishPoolParameters}
\end{table}

%%%%%%%%%%%%%%%%%%%%%%%%%%%%%%%%%%%%%%%%%%%%%%%%%%%%%%%%%%%%%%%%%%%%%%%%%%%%%%%%%
%%% 
%%%%%%%%%%%%%%%%%%%%%%%%%%%%%%%%%%%%%%%%%%%%%%%%%%%%%%%%%%%%%%%%%%%%%%%%%%%%%%%%%
\subsection{Optimal Stopping Problem}\label{sec:stopping}
Following \citet{EK2023} the objective of a fish farmer is to find the optimal harvesting time of the salmon cultivated in the farm, where optimal has to be understood as obtaining the maximal expected value under a risk-neutral measure.
This can be formalized by considering the current value of the fish at their current weight $S_t^1 \biomass(t)$ minus the harvesting costs at $\cumHC(t)$ and the cumulative feeding costs up to this point in time $\cumFC(t)$. In the case of deterministic feeding costs and the host-parasite model, let $X_t^\text{stoch}\coloneqq \left(S_t^1,\delta_t^1,H_t,P_t\right)$ and $X_t^\text{determ}\coloneqq \left(S_t^1,\delta_t^1\right)$. In the case of stochastic feeding costs $\left(S_t^2,\delta_t^2\right)$ needs to be added. Thus, the dimension of $X_t$ is $d=6$ in the stochastic host-parasite model with stochastic feeding costs and less for all the other cases.

For a fixed choice of feeding model, the optimal stopping problem for stochastic and deterministic mortality becomes respectively
\begin{align*}
    W_0^\text{stoch}(x)&\coloneqq
    \sup_{\tau^\text{stoch}}
        \E^\Q\left[\left.
            \exp\left(- r \tau\right) 
            \left(
               \left(1-\cumTreatmentCosts_t^\text{stoch} \right) S_\tau^1 \biomass_\tau - \cumHC_\tau
            \right)
             - \cumFC_\tau
             \right|
             X_0^\text{stoch}=x
        \right],\\
    W_0^\text{determ}(x)&\coloneqq
    \sup_{\tau^\text{determ}}
        \E^\Q\left[\left.
            \exp\left(- r \tau\right) 
            \left(
               \left(1-\cumTreatmentCosts_t^\text{determ} \right) S_\tau^1 \biomass_\tau - \cumHC_\tau
            \right)
             - \cumFC_\tau
             \right|
             X_0^\text{determ}=x
        \right].
        %\label{eq:optStopping}
\end{align*}
% \todo{Choice of measure}
In this paper, we compare the stopping rule obtained from $W_0^\text{stoch}(x)$ and $W_0^\text{determ}(x)$ by evaluating
\begin{align}
 V_0^z(x)\coloneqq \E^\Q\left[\left.
    \exp\left(- r \tau^z\right) 
    \left(
        \left(1-\cumTreatmentCosts_t^\text{stoch} \right) S_{\tau^z}^1 \biomass_{\tau^z} - \cumHC_{\tau^z}
    \right)
     - \cumFC_{\tau^z}
     \right|
     X_0^\text{stoch}=x
     \label{eq:objective}
\right]   
\end{align}
for $z=\text{stoch},\,\text{determ}$ in an appropriate way described in \Cref{sec:stoppingBoundaries}, and answer the question whether $V_0^\text{determ}(x)$ is a good approximation of $W_0^\text{stoch}(x)$
or if it is beneficial to consider the slightly more complicated stopping rule $\tau^\text{stoch}$.

We will solve the optimal stopping problems numerically by using the Deep Optimal Stopping Network by \citet{Becker2021} and refer for the algorithmic details to the same paper. A simple least-square Monte Carlo (LSMC) approach with a polynomial basis, led to numerically ill-conditioned matrices for solving the regression step and inaccurate results. We leave it to future research to find a more suitable basis, e.g., a neural network.

%%%%%%%%%%%%%%%%%%%%%%%%%%%%%%%%%%%%%%%%%%%%%%%%%%%%%%%%%%%%%%%%%%%%%%%%%%%%%%%%%
%%% 
%%%%%%%%%%%%%%%%%%%%%%%%%%%%%%%%%%%%%%%%%%%%%%%%%%%%%%%%%%%%%%%%%%%%%%%%%%%%%%%%%
\section{Numerical Experiments}\label{sec:numerics}
In this section, we will perform our numerical experiments. In \Cref{sec:marketData}, we will briefly discuss the historical data used for the commodity models, as well as fix their parameters. This is followed by a detailed examination of the available salmon lice data in \Cref{sec:liceData}, which will be used to calibrate the host-parasite model in \Cref{sec:calibration}. After the calibration, we will compare the stochastic mortality model to its deterministic counterpart in \Cref{sec:stoppingBoundaries} and reach our conclusion that it can be beneficial to consider stochastic mortality models.

%%%%%%%%%%%%%%%%%%%%%%%%%%%%%%%%%%%%%%%%%%%%%%%%%%%%%%%%%%%%%%%%%%%%%%%%%%%%%%%%%%%%%%%%%%%
%% Hardware and Software configuration
For the calibration we used \matlab with the \matlabGOtoolbox
and for the DNN \python with \tensorflow
running on \OS, on a machine with the following specifications: processor
\CPU and \RAM. A GPU did significantly improve the performance in this case and we 
used an \GPU.

For all of our experiments, we will fix the interest rate to $r=0.0303$.
%%%%%%%%%%%%%%%%%%%%%%%%%%%%%%%%%%%%%%%%%%%%%%%%%%%%%%%%%%%%%%%%%%%%%%%%%%%%%%%%%
%%% 
%%%%%%%%%%%%%%%%%%%%%%%%%%%%%%%%%%%%%%%%%%%%%%%%%%%%%%%%%%%%%%%%%%%%%%%%%%%%%%%%%
\subsection{Future Market Data}\label{sec:marketData}
We use the same historical future data \fromDate till \toDate for both salmon and soy as in \citet{EK2023} and chose the following parameters that had been obtained there:
\begin{align}
    \begin{aligned}
        \sigma_1^1=0.23 ,\
        \sigma_2^1=0.75 ,\
        \kappa^1=2.6 ,\
        \alpha^1= 0.02,\
        \lambda^1=0.2 ,\
        \rho^1= 0.9,\
        \delta^1_0=0.57 ,\
        \hat{S}_0^1= 95,
        \\
        \sigma_1^2=1 ,\
        \sigma_2^2=0.4 ,\
        \kappa^2= 1.2,\
        \alpha^2= 0.06,\
        \lambda^2= 0.14,\
        \rho^2= 0.44,\
        \delta^2_0= 0,\
        \hat{S}_0^2= 1500.
    \end{aligned}
    \label{eq:soySalmonParam}
\end{align}

%%%%%%%%%%%%%%%%%%%%%%%%%%%%%%%%%%%%%%%%%%%%%%%%%%%%%%%%%%%%%%%%%%%%%%%%%%%%%%%%%
%%% 
%%%%%%%%%%%%%%%%%%%%%%%%%%%%%%%%%%%%%%%%%%%%%%%%%%%%%%%%%%%%%%%%%%%%%%%%%%%%%%%%%
\subsection{Lice and Treatment Data}\label{sec:liceData}
In this section, we explain what data is available for estimating stochastic mortality and treatment costs.\footnote{The data we used is publicly available at \url{https://www.barentswatch.no/nedlasting/fishhealth/lice} (last accessed 31/07/2023 13:04 CEST).} 
The data contains information about salmon lice per fish, moving lice per fish, and stuck lice per fish, as well as removal and treatment measures to counteract the lice infestation for each fishing site in Norway. These counts are updated weekly since \fromDate and are available till today.

We are aware, that the lice growth is affected by the salinity and temperature of the water and therefore restrict ourselves to farms in a specific region of Norway, namely Tr\o ndelag, to avoid dealing with immense fluctuations of those.

\begin{figure}
    \centering
    \includegraphics[width=\columnwidth]{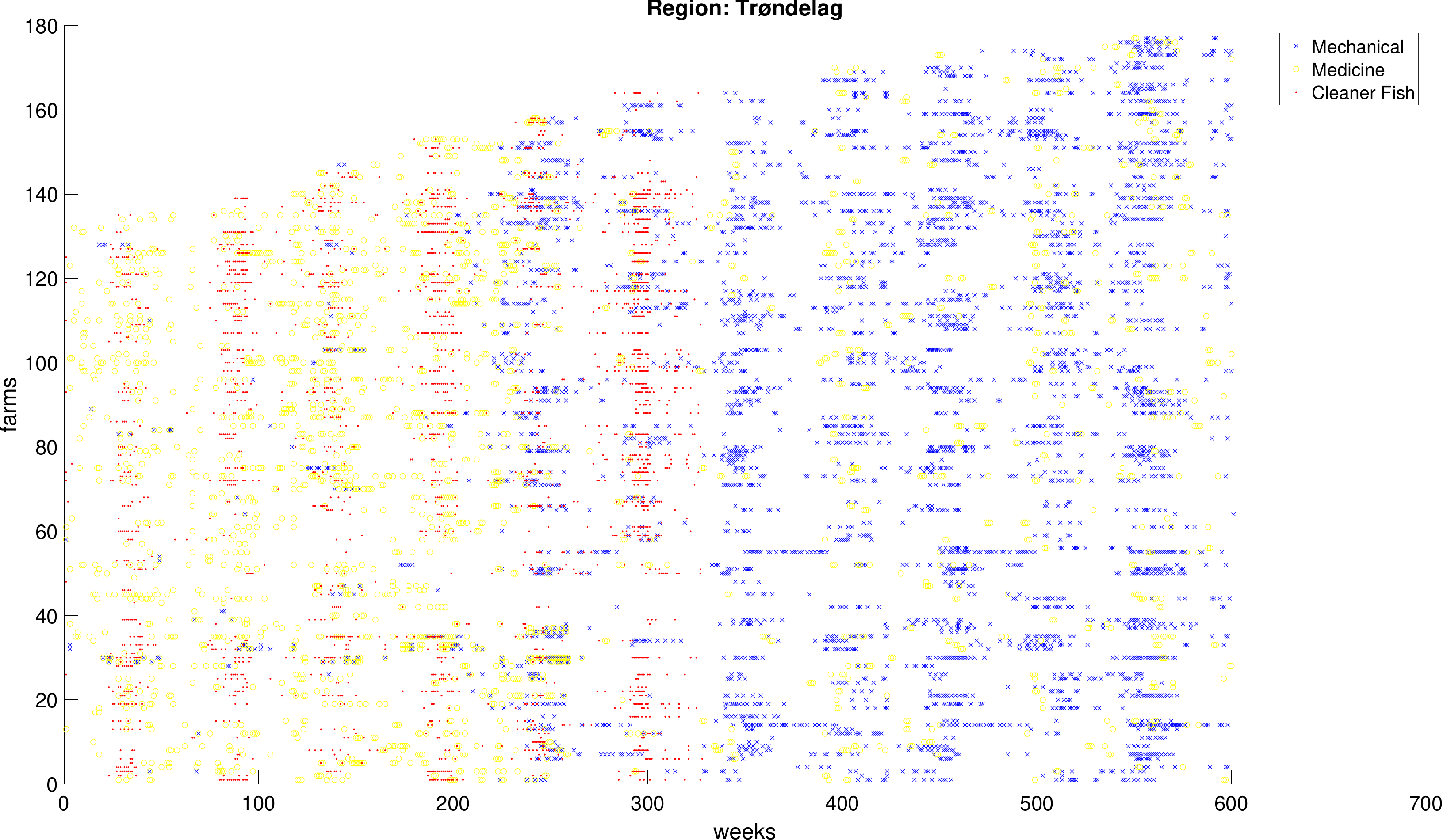}
    \caption{Evolution of treatment techniques in Tr\o ndelag from 2012 till today.}
    \label{fig:treatments}
\end{figure}

There are three major categories of lice removals: Cleaner fish, medical treatments, and mechanical removals. 
In \Cref{fig:treatments}, we can see the chosen treatments and treatment times depicted by colored marks of roughly 200 fish farms. 
The usage of cleaner fish (red dots) was very popular from 2012 till roughly 2018 but is not used anymore in more recent years. Also, the usage of medical treatments (yellow dots) seems to have lessened over time. On the other hand, mechanical removals (blue dots) of lice have become the method of choice in more recent years and will be the focus of lice countermeasures in this paper. To give an idea of what mechanical removal refers to, there are methods using brushes to brush the lice of the salmon skin, hot-water or fresh-water baths, as well as laser techniques. We will not distinguish between different methods and note that each of these methods has a very small probability of harming or killing the salmon in the process with varying effectiveness of removing the lice.

\begin{figure}
    \centering
    \includegraphics[width=\columnwidth]{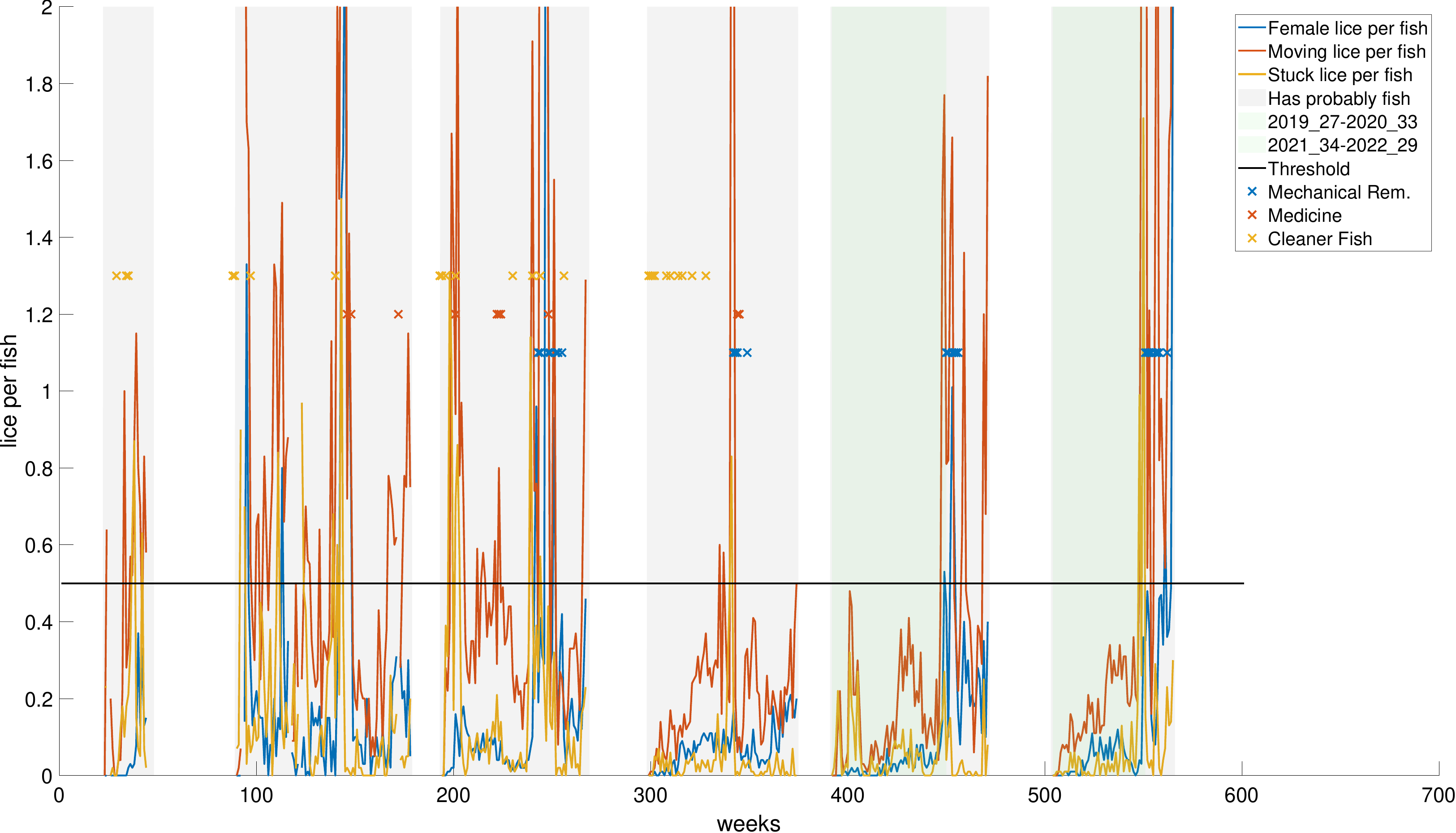}
    \caption{Evolution of lice per fish, as well as countermeasures for one specific salmon farm from 2012 till today.}
    \label{fig:lpfFarm}
\end{figure}

In \Cref{fig:lpfFarm}, we selected a random fish farm in Tr\o ndelag region, to show the evolution of female lice per fish (blue bold line), 
moving lice per fish (red bold line), stuck lice per fish (yellow bold line), as well as the times of using mechanical removals, medical treatments, or cleaner fish depicted with blue, red, and yellow crosses respectively.
The light-gray areas show the farm periods from planting smolt (young salmon) to the harvesting of the farm. 
The bold black line marks the maximal female lice per fish threshold, which must not be exceeded for a longer period by Norwegian regulations of fish farms. If a farmer would ignore this threshold or would use insufficient countermeasures, the farm would be shut down and all fish would need to be exterminated to prevent a salmon lice epidemic in the area.

We can see that the lice per fish can behave very differently in these periods if we compare the earlier periods with the later ones, which may or may not be due to the different treatments. Thus, for our investigation, we restrict ourselves to datasets only having mechanical removals and no other treatments, like in the last two periods. The green areas mark the sections of the periods till the first mechanical removal. We will use these green sections to estimate the reproduction rate of salmon lice in the host-parasite model.

\begin{figure}
    \centering
    \includegraphics[width=\columnwidth]{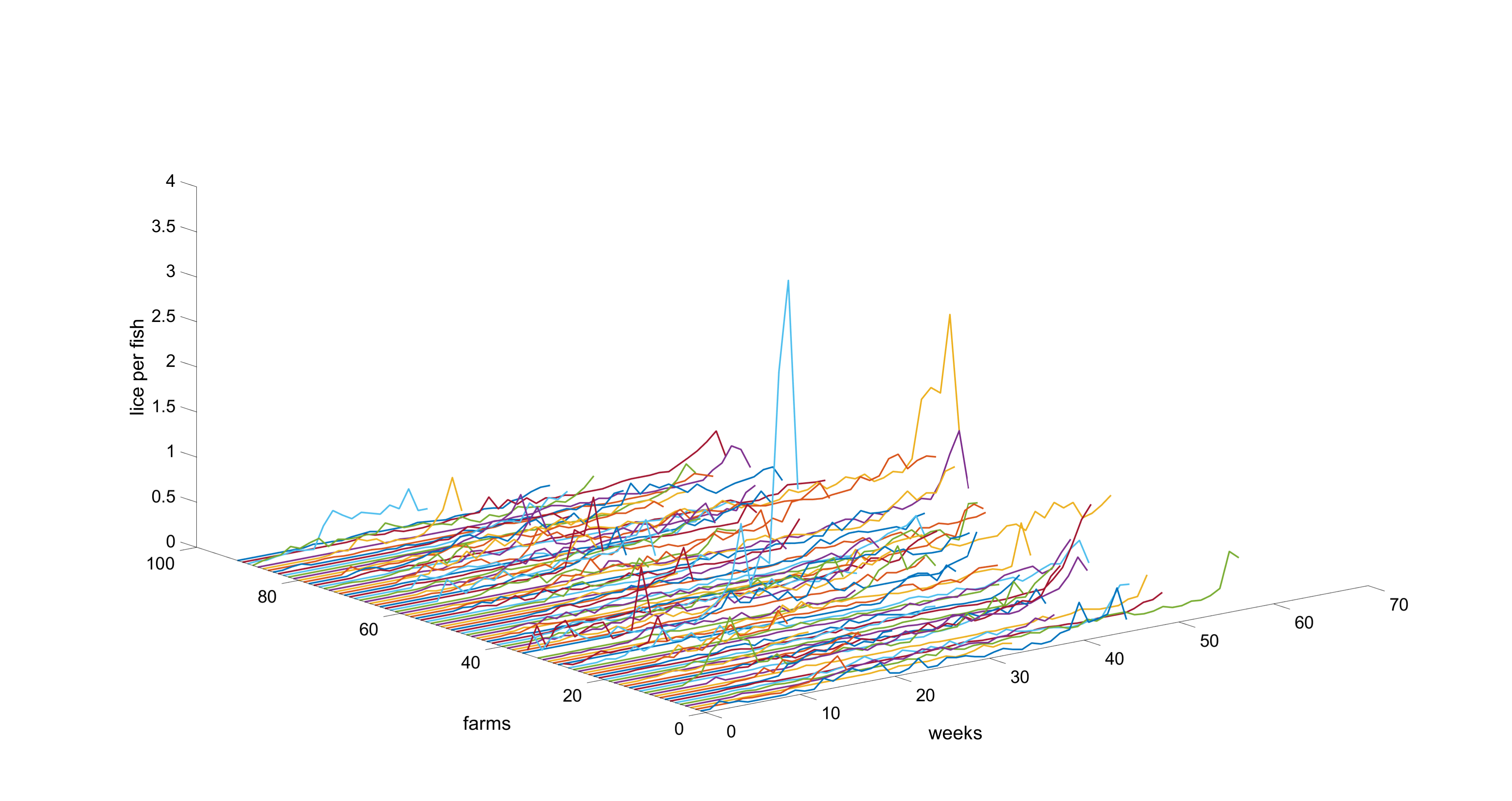}
    \caption{Evolution of lice per fish till first mechanical removal for farms in Tr\o ndelag, excluding all periods with additional treatments to mechanical removals.}
    \label{fig:lpfRegion}
\end{figure}

In \Cref{fig:lpfRegion}, we show the roughly 100 selected green areas of all of the fish farms in Tr\o ndelag region. Each trajectory (in various colors) is one evolution of female salmon lice per fish from the start of the fish farm till a varying treatment time. There is no immediate clear pattern recognizable, except for an exponential growth with varying rates. This can be due to various factors, for example, preventive removals, i.e., not waiting till the threshold is exceeded or close to it, a different number of lice per fish (in different stages of their life cycles) at the start of the farm, different rate of lice coming from the ocean, human error, etc. Since we are no experts in this field, we will treat this as noise.

\begin{figure}
    \centering
    \includegraphics[width=\columnwidth]{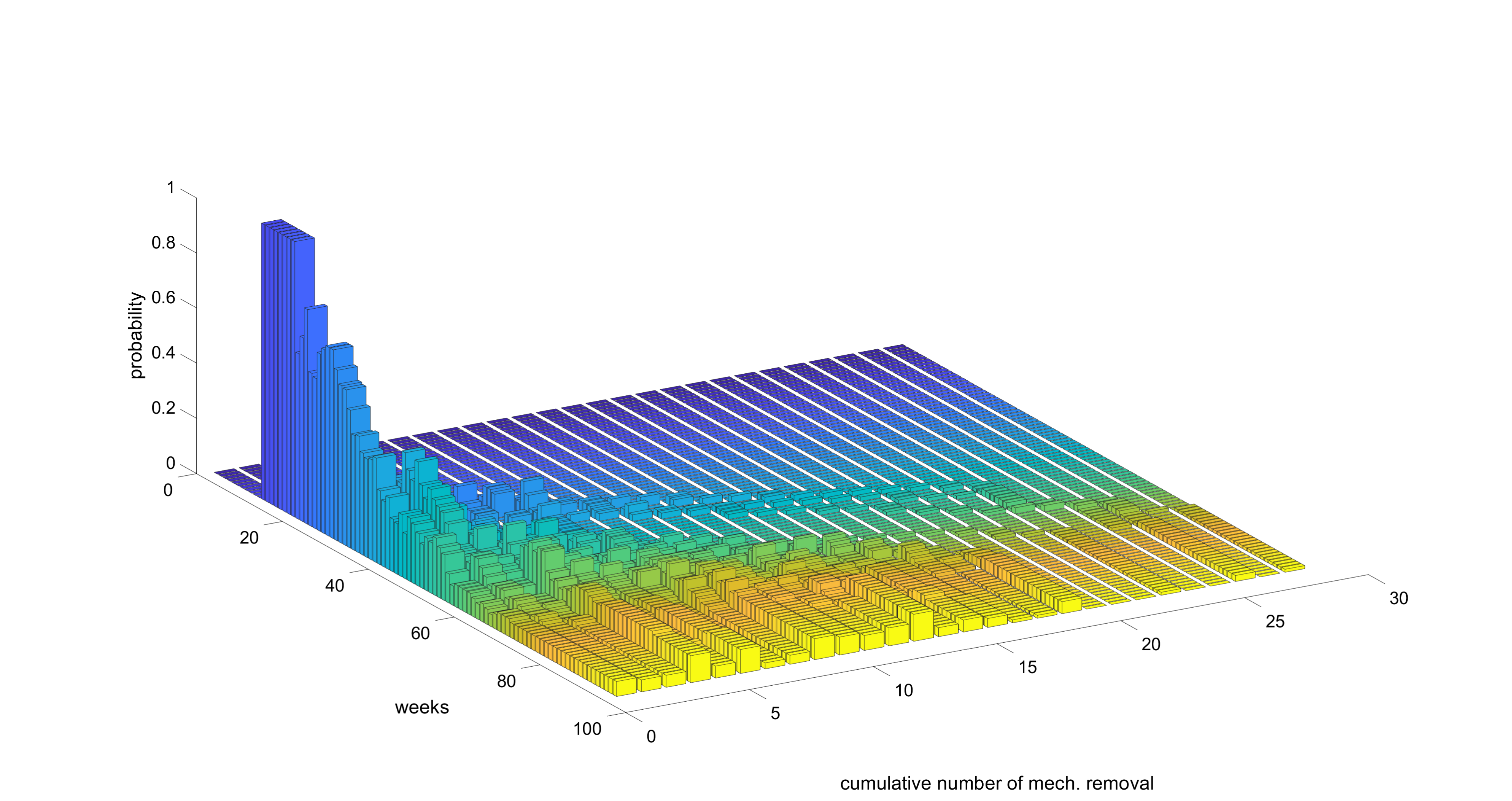}
    \caption{Evolution of distributions of cumulative mechanical removals for farms in Tr\o ndelag, excluding all periods with additional treatments to mechanical removals.}
    \label{fig:evoRemovals}
\end{figure}

Now, that we have a better understanding of the evolution of lice, we focus in \Cref{fig:evoRemovals} on the distribution of accumulated mechanical removals from the start till harvesting the fish farms. On the x-axis, we see the time in weeks, starting in the back at zero and going up to 100 weeks in the front of the picture. The y-axis shows the number of accumulated mechanical removals over time and the z-axis the corresponding probability of this number. Hence, each line (in the y-axis) for a fixed week is a probability distribution of the accumulated number of removals over time. We see that in the first 10 weeks, there is no treatment at all, then the entire mass is focused on one treatment at roughly week 15, which spreads out afterwards. From week 70 onward (the yellowish area), the distribution is quite stable and is because harvesting will occur for the majority of the farms in this period, making further treatments close to the harvesting time unnecessary. The mean of treatments in the yellow area is roughly 10. We assumed that the total number of treatment costs is roughly 30\,\% (see \Cref{tab:fishPoolParameters}) of the production costs, which is 50\,\% of the salmon value, and therefore 15\,\% of the overall value.
This means that $\treatmentCosts=1.5\,\%$ is a good approximation for each treatment.

%%%%%%%%%%%%%%%%%%%%%%%%%%%%%%%%%%%%%%%%%%%%%%%%%%%%%%%%%%%%%%%%%%%%%%%%%%%%%%%%%
%%% 
%%%%%%%%%%%%%%%%%%%%%%%%%%%%%%%%%%%%%%%%%%%%%%%%%%%%%%%%%%%%%%%%%%%%%%%%%%%%%%%%%
\subsection{Calibration of the Host-Parasite Model}\label{sec:calibration}
In this section, we will explain how to calibrate the host-parasite model introduced in \Cref{sec:stochMortality} to our selection of the data, i.e., only farming periods in Tr\o ndelag region, where only mechanical removals were used as a lice treatment leading to roughly 100 different farming periods.

We partition the calibration of the host-parasite model \eqref{eq:hostParasite} into two steps: First, we fit the deterministic part to the data shown in \Cref{fig:lpfRegion}, i.e., the data of female lice per fish till the first mechanical removal, to infer the reproduction rate of the parasites. Note that we will disregard for simplicity the different phases that the copepodites undergo through their life cycle and thus also the data on stuck lice per fish and moving lice per fish. After inferring the reproduction rate of the salmon lice, we find the parameters for the effectiveness $Y_t\sim \mathrm{beta}(\beta_1,\beta_2)$ to match the distribution of cumulative mechanical removals in the yellow area of \Cref{fig:evoRemovals}.

Henceforth, let us fix the intrinsic mortality of salmon to $\mu=0.05$, the mortality due to salmon lice infestation $\alpha=0.1$, and the intrinsic mortality of salmon lice $b=0.05$. We found these parameters reasonable without any biological validation at this point and encourage the reader to change these values to their preferences in the code. As aforementioned, we will infer in the first step the reproduction rate $\lambda$ from the data. Thus, we want to solve the following least-squares problem
\begin{align*}
    \min_{\lambda>0}
    \sum_{k=1}^{K} \sum_{i=1}^{N} \left(\frac{P_{t_i}}{H_{t_i}}-\LPF^k_{t_i} \right)^2,
\end{align*}
where $\LPF^k_{t_i}$ are the female lice per fish of the $k$-th farm (y-axis in \Cref{fig:lpfRegion}) at time $t_i$ (x-axis in \Cref{fig:lpfRegion}). We solve the coupled ODE for $H_t$ and $P_t$ by an explicit Euler-scheme with constant step size $\Delta t = \frac{T}{10N -1}$ ($T$ and $N$ as in \Cref{tab:fishPoolParameters}). Notice, that $H_t$ and $P_t$ are deterministic till the first passing time $\frac{P_t}{H_t}>\liceThreshold(t)$. 

We found for the initial number of salmon $H_0=10000$ and $0.1\,\%$ lice per fish the reproduction rate $\lambda=7.0143$.

In a second step, we simulate $M=1000$ trajectories of the host-parasite model with jumps at the hitting times 
$\frac{P_t}{H_t}=\liceThreshold(t)$ with parameters $\beta_1,\beta_2$ for the effectiveness of the treatments. Hence, we want to match the first two moments at $t=1.77$ (yellow area in \Cref{fig:evoRemovals}), i.e.
\begin{align*}
    \min_{\beta_1>0,\beta_2>0}
    \left(\E\left[\sum_{0\leq s \leq t}{\Delta H_s}\right] -\mathrm{Mean}\left[\cumRemovals_t\right]\right)^2+
    \zeta\left(\mathrm{std}\left[\sum_{0\leq s \leq t}{\Delta H_s}\right] -\mathrm{std}\left[\cumRemovals_t\right]\right)^2,
\end{align*}
where $\zeta>0$ is a parameter to adjust the emphasis on the standard deviation during the calibration\footnote{In our experiments, we found that $\zeta=2$ works best and use this choice in all of the following tests.} and $\cumRemovals_t$ denotes the vector of cumulative mechanical removals up to time $t$. We found the parameters $\beta_1=0.0829$ and $\beta_2=0.0281$. 

We used \matlab's \lsqnonlin with the trust-region-reflective algorithm for the first optimization problem, which took 0.1 seconds, and \fmincon with the interior point method for the second one, taking 19.8 seconds on CPU.

\begin{figure}
    \centering
    \includegraphics[width=\columnwidth]{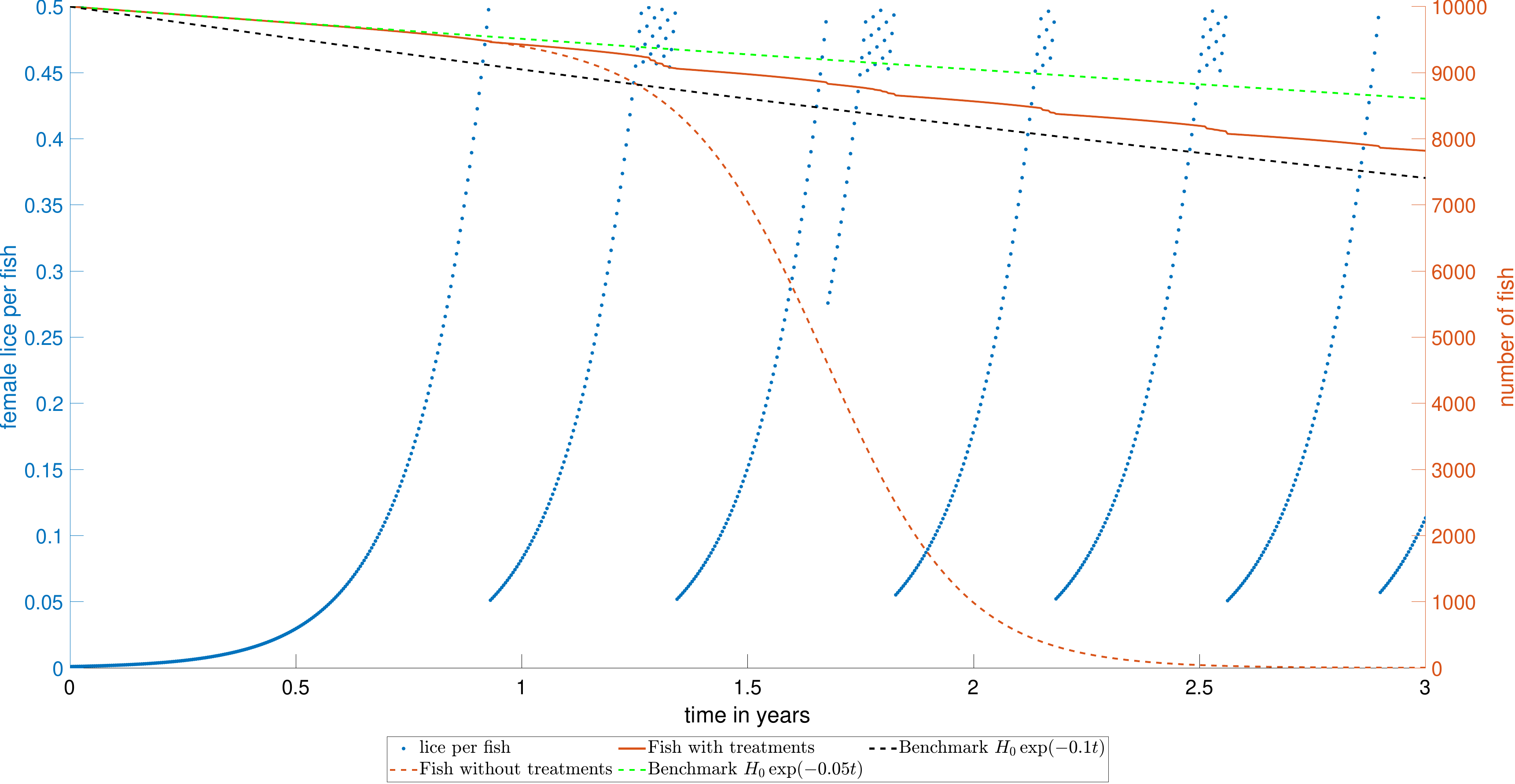}
    \caption{One trajectory of the simulated host model with treatments (bold red line) and corresponding lice per fish in blue dots.}
    \label{fig:hostParasiteTra}
\end{figure}

In \Cref{fig:hostParasiteTra}, we show one trajectory of the simulated host (red bold line) and parasite population (dotted blue lines), as well as what would happen to the salmon (red dashed line), if no lice-removals were applied in our model.
The left y-axis corresponding to the blue dots shows the lice per fish and the right y-axis corresponds to the number of salmon. The x-axis shows the time-axis in years. Let us focus on the red lines first. We can see that till roughly $t=0.75$ they are the same because the model is deterministic till the first hitting time. After $t=0.75$ the dotted line shows what would happen if the lice were not removed, the salmon population would quickly decline in this model. We also show two benchmarks in green and black dashed lines with deterministic mortality rates of $\moratlityRate=0.05$ and $\moratlityRate=0.1$, respectively.
By the distance of the green and red dotted line, we can see the impact of the increasing lice population. In this model, we can see that as long as the lice per fish is below the threshold $0.5$, the mortality due to lice has no significant impact on the salmon, which should reflect the reality well. Now, let us focus on the lice per fish, i.e., the blue dots. We can see the jump times well by the sudden drops, followed by an exponential increase afterwards till the lice threshold is exceeded. The height of the jumps is determined by the random numbers $Y_t$ drawn from the calibrated beta distribution. At each of these jumps a uniform random number $X_t$ is drawn as well to reflect the small probability of salmon dying from lice removal treatments, which explains the jumps in the bold red line.

In \Cref{fig:hostParasiteTra}, we show histograms illustrating the distribution of cumulative mechanical removals at two specific points in time
$t=1.09$ on the left-hand side and $t=1.77$ on the right-hand side. The blue bars correspond to the simulated host-parasite model and the red bars to the data (slices of \Cref{fig:evoRemovals}).

\begin{figure}
    \centering
    \includegraphics[width=.49\columnwidth]{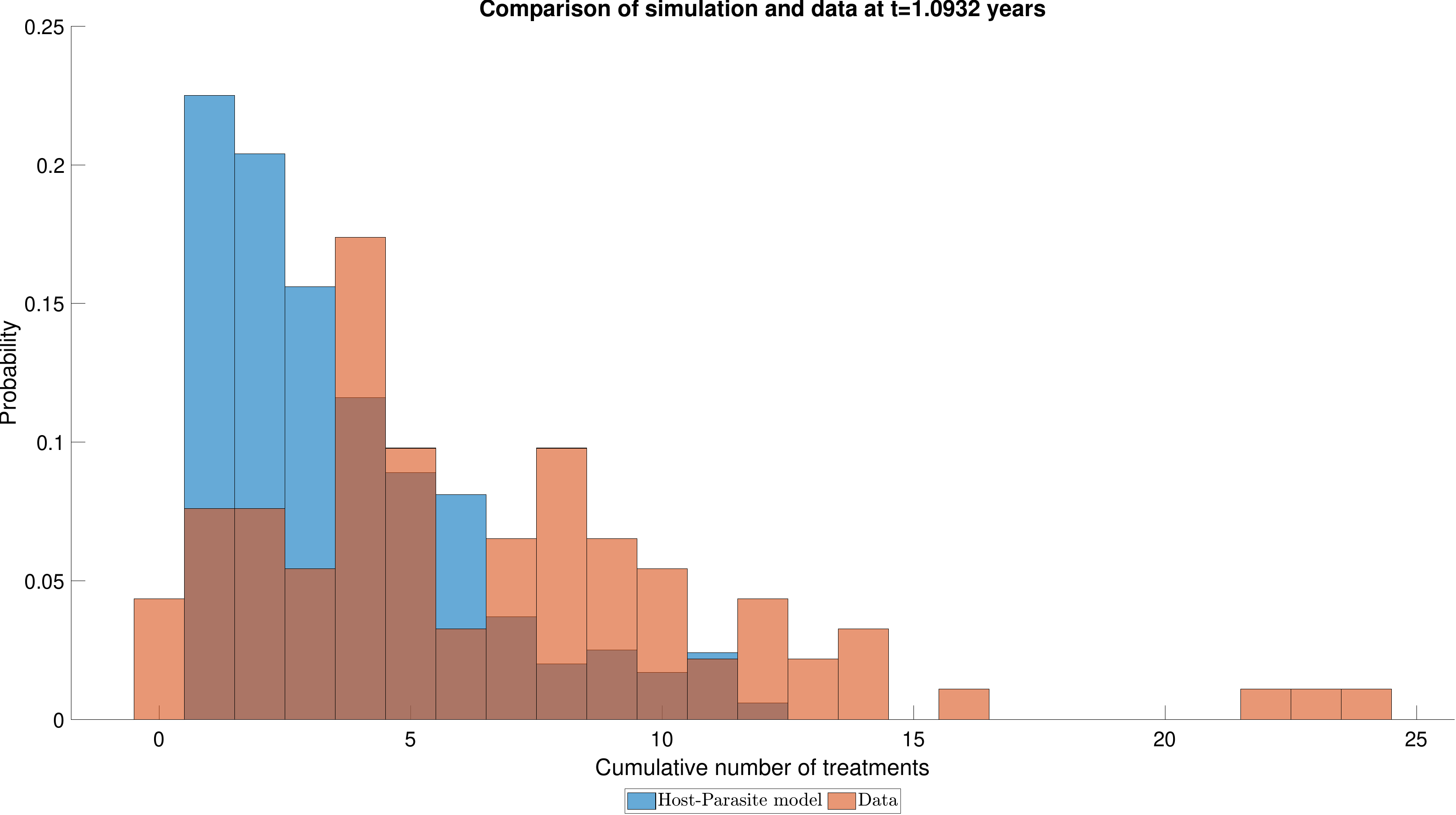}
    \includegraphics[width=.49\columnwidth]{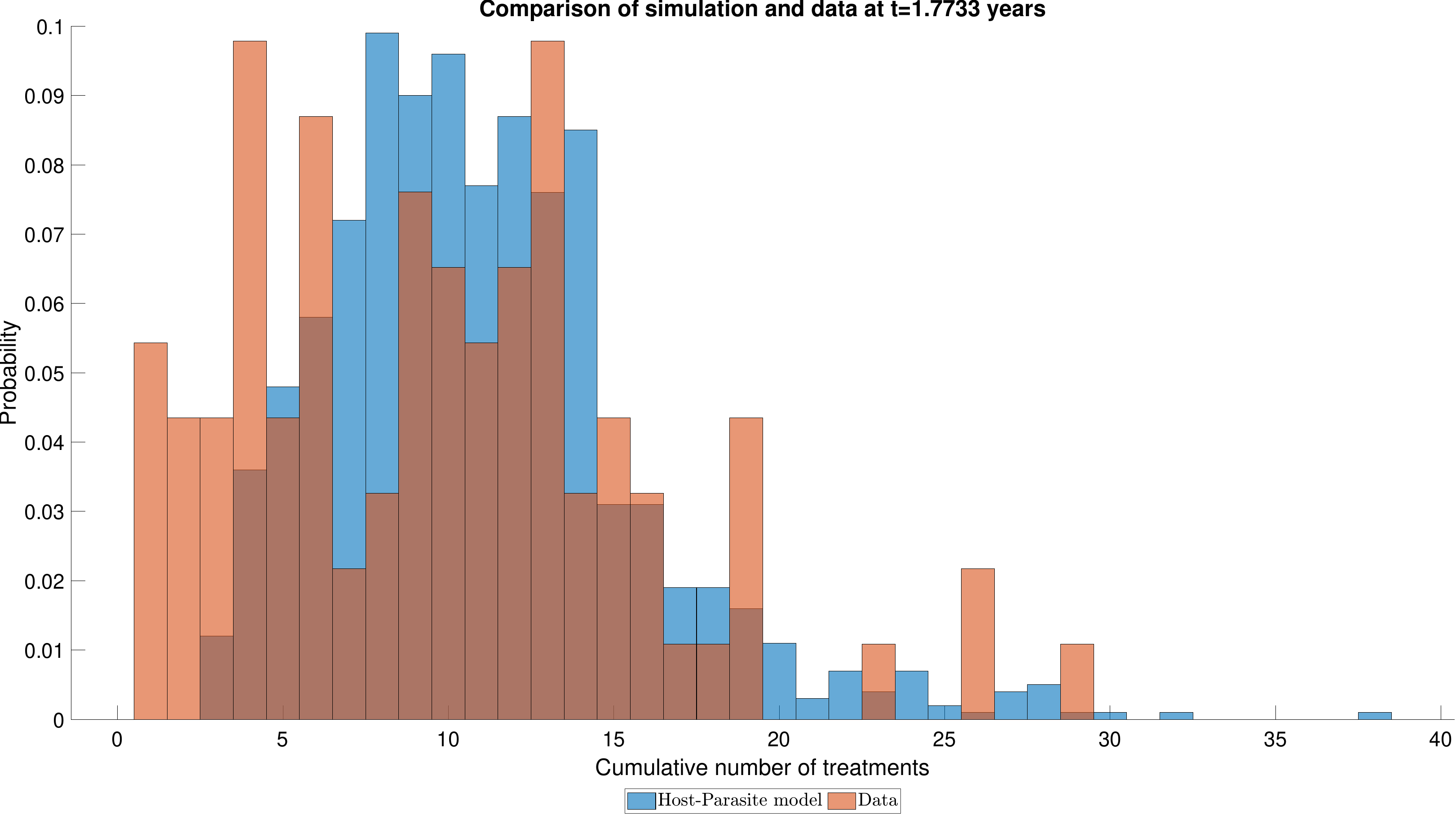}
    \caption{Distribution of the number of cumulative mechanical removals using the host-parasite model (blue bars). The left figure compares the model distribution to the distribution from the data at $t=1.09$ and the right figure at $t=1.77$.}
    \label{fig:hostParasiteDist}
\end{figure}

We can see that the raw data exhibits two modes, while our model has a skewed distribution with one mode. At $t=1.77$ they match well enough in our opinion, since we can expect a lot of noise in this small dataset. For $t=1.09$ a smaller number of mechanical removals is estimated by our model than seen from the data. Including more points in time for the second calibration problem could average these two fits, but we found the model distributions reasonable for our further investigations.

%%%%%%%%%%%%%%%%%%%%%%%%%%%%%%%%%%%%%%%%%%%%%%%%%%%%%%%%%%%%%%%%%%%%%%%%%%%%%%%%%
%%% 
%%%%%%%%%%%%%%%%%%%%%%%%%%%%%%%%%%%%%%%%%%%%%%%%%%%%%%%%%%%%%%%%%%%%%%%%%%%%%%%%%
\subsection{Comparison of Exercise Decisions}\label{sec:stoppingBoundaries}
In \citet{EK2023}, we found that the difference in exercise decisions can be estimated from a pathwise comparison of the stopping times. It led to the same conclusions as determining the decision boundary first, so we will just use the pathwise comparison here, which we will recall for the reader's convenience:

Let $\left(\Omega,\mathcal{F},\mathcal{F}_t,\P\right)$ be a filtered probability space satisfying the usual conditions, where
$\mathcal{F}_t\coloneqq \sigma\left(X_s\vert s\leq t\right)$ is generated by a stochastic process, and
$\mathcal{F}^i_t \subseteq \mathcal{F}_t$, $i=1,2$, be two sub-filtrations.
Moreover, let $\tau^1$ be an ($\mathcal{F}^1_t$-)stopping time and
$\tau^2$ an ($\mathcal{F}^2_t$-)stopping time.
Now, let us fix a path $(X_t(\omega))_{t\geq 0}$, then conditioned on this path the two stopping times $\tau^1(\omega)\in [0,T]$ and $\tau^2(\omega)\in [0,T]$ can be compared as real-numbers.

Additionally, still conditioned on $(X_t(\omega))_{t\geq 0}$, we may compare the stopped values of a stochastic process $Y_t$, i.e.,
$Y_{\tau^1(\omega)}$ and $Y_{\tau^2(\omega)}$.

We will compare now the impact of the stochastic mortality by comparing $V_0^{\text{stoch}}$ and $V_0^{\text{determ}}$ using the stopping times obtained from $W_0^{\text{stoch}}$ and $W_0^{\text{determ}}$ in the host-parasite model. 

As aforementioned, we solve the optimal stopping problem by using the DeepOS network by \citet{Becker2021}, since in the standard LSMC approach the polynomial basis led to poorly conditioned matrices for solving the linear systems, a ridged regression did not help in this case. Using neural networks as basis functions might improve the performance and we leave this for future considerations. The training of the Deep Neural Networks with $1500+d$ epochs and a batch-size of $2^{12}$ is quite fast using a GPU, it takes in the fully stochastic case roughly 90 seconds using the host-parasite model and is a bit faster in the other cases. The majority of the computational time comes from the simulation of the stochastic processes. For the evaluation, we use $20\cdot 2^{12}$ trajectories and the evaluation of the neural networks takes less than a second.

In \Cref{tab:relImprovementsStoch}, we show the relative improvement of using a stochastic mortality model compared to deterministic mortality, denoted by $\relImprovement\coloneqq \frac{V_0^{\tau,\text{stoch}}}{V_0^{\tau,\text{determ}}}$. Let us focus on the scenario with stochastic feeding costs first, i.e. the second column.
We can see that the mean stopping times are quite similar and the relative improvement is roughly $1.5\,\%$. 

\begin{table}[]
    \caption{Relative improvements using a stochastic mortality model with deterministic and stochastic feeding costs.}
    \centering
    \begin{tabular}{l*{2}{c}}
                                    & Stochastic feeding costs  & Deterministic feeding costs  \\
        \toprule
        $\E[\tau^{\text{stoch}}]$   & 1.989             & 2.006  \\
        $\E[\tau^{\text{determ}}]$  & 1.993             & 2.005\\
        $V_0^{\text{stoch}}$        & 2501648           & 2442247\\
        $V_0^{\text{determ}}$       & 2466925           & 2406204\\
        $\relImprovement$           & 1.014             & 1.015\\
    \end{tabular}
    \label{tab:relImprovementsStoch}
\end{table}

In the scenario with deterministic feeding costs, we see similar results. Therefore, it seems like the benefit of using a stochastic mortality model is independent of the chosen feeding cost model since the relative improvements changed only slightly between the stochastic and deterministic feeding cost models.

Changing $\treatmentCosts$ to $1\,\%$ lessens the improvements to roughly $0.7\,\%$ in the host-parasite model and 
increases the improvements to $2.3\,\%$ when changing $\treatmentCosts$ to $2\,\%$.
%%%%%%%%%%%%%%%%%%%%%%%%%%%%%%%%%%%%%%%%%%%%%%%%%%%%%%%%%%%%%%%%%%%%%%%%%%%%%%%%%
%%% 
%%%%%%%%%%%%%%%%%%%%%%%%%%%%%%%%%%%%%%%%%%%%%%%%%%%%%%%%%%%%%%%%%%%%%%%%%%%%%%%%%
\section{Conclusion and Limitations}\label{sec:conclusion}
In this article, we focused on the question of whether stochastic mortality models lead to a significant increase in revenue compared to deterministic mortality models.
In the case of a simple host-parasite model, we found an affirmative answer leading to an increase of $1.5\,\%$ of the farm's value. In future studies, we would like to find a more realistic model for the host-parasite relationship, taking the different phases of the copepodites into account and therefore using the entire dataset. Also, a biological validation of the model and its parameters is required to judge how realistic this mathematical model will be. Another economical validation will be required to judge how the treatment costs are incorporated into the optimal stopping problem. 

Another idea for future research would involve the consideration of an optimal treatment strategy instead of just ensuring that the lice threshold is not exceeded. Maybe preventive mechanical removals in earlier stages could be beneficial. Moreover, including medical treatments with stochastic delay differential equations (SDDE) could be of interest.
%%%%%%%%%%%%%%%%%%%%%%%%%%%%%%%%%%%%%%%%%%%%%%%%%%%%%%%%%%%%%%%%%%%%%%%%%%%%%%%%%%%
%%% Appendix
%%%%%%%%%%%%%%%%%%%%%%%%%%%%%%%%%%%%%%%%%%%%%%%%%%%%%%%%%%%%%%%%%%%%%%%%%%%%%%%%%%%
\appendix

%%%%%%%%%%%%%%%%%%%%%%%%%%%%%%%%%%%%%%%%%%%%%%%%%%%%%%%%%%%%%%%%%%%%%%%%%%%%%%%%%%%%%%%%%%%
%% Declarations
%%%%%%%%%%%%%%%%%%%%%%%%%%%%%%%%%%%%%%%%%%%%%%%%%%%%%%%%%%%%%%%%%%%%%%%%%%%%%%%%%%%%%%%%%%%
\section*{Declarations}
The authors have no relevant financial or non-financial interests to disclose.

%\nocite{*}
\printbibliography

\end{document}